\documentclass[conference]{IEEEtran}
\usepackage{cite}
\usepackage{amsmath,amssymb,amsfonts}
\usepackage{algorithmic}
\usepackage[bookmarks=false]{hyperref}
\usepackage{listings}
\usepackage{graphicx}
\usepackage{textcomp} 
\usepackage{xcolor}
\usepackage{tabularx}
\usepackage{array}
\usepackage{booktabs}
\usepackage{enumitem}
\usepackage{mathrsfs}
\usepackage{wasysym}
\usepackage{makecell} 

\usepackage{xcolor} 
\definecolor{light-gray}{gray}{0.95}

\begin{document}

\title{Model-Based Engineering of CPPS Functions and Code Generation for Skills}

\author{
\IEEEauthorblockN{
    Aljosha Köcher\IEEEauthorrefmark{1},
    Alexander Hayward\IEEEauthorrefmark{1},
    Alexander Fay\IEEEauthorrefmark{1},
}
\IEEEauthorblockA{
\IEEEauthorrefmark{1}Institute of Automation\\
Helmut Schmidt University, Hamburg, Germany\\
Email: aljosha.koecher@hsu-hh.de, alexander.hayward@hsu-hh.de, alexander.fay@hsu-hh.de\\}
}

\maketitle

\begin{abstract}
Today's production systems are complex networks of cyber-physical systems which combine mechanical and electronic parts with software and networking capabilities. To the inherent complexity of such systems, additional complexity arises from the context in which these systems operate. Manufacturing companies need to be able to adapt their production to ever changing customer demands as well as decreasing lot sizes.
Engineering such systems, which need to be combined and reconfigured into different networks under changing conditions, requires engineering methods to carefully design them for possible future uses.
Such engineering methods need to preserve the flexibility of functions into runtime, so that reconfiguring machines can be done with as little effort as possible. 
In this paper we present a model-based approach that is focused on machine functions and allows to methodically develop system functionalities for changing system networks. These functions are implemented as so-called skills using automated code-generation.
\end{abstract}

\begin{IEEEkeywords}
MBSE, Functions, Functional Architecture, Skills, Code Generation
\end{IEEEkeywords}

\section{Introduction}
Today's manufacturing systems are deployed in environments that are shifting with increasing frequency and speed. One of these shifts are regularly changing customer requirements due to changing product demands. A cyber-physical production system (CPPS) is described as independent modules equipped with different capabilities to contribute to the production of individual customer orders. To this purpose, modules organize themselves into temporary, dynamic manufacturing networks to enable manufacturing products by combining their different capabilities. Modules are developed by different vendors, who have to anticipate and consider collaboration in CPPS and their changing constitution already at design time.\cite{SchlingelholgerCrestBuch}

The complexity of information to be considered during design time has increased steadily in recent decades due to the growing number of system functionalities \cite{Vogelsang.2015}. Common methods for managing this complexity recommend approaches of model-based systems engineering (MBSE) that describe the individual system using several levels of granularity, from requirements to functions and lastly components to be implemented. 
In order to enable an individual module to collaborate in a CPPS, corresponding information about the behavior of the individual systems within a CPPS must be specified \cite{SPESXT}. 
Accordingly, a manufacturer-independent model of the CPPS would be ideal, from which all the requirements for the individual system can be derived.
From the perspective of each individual manufacturer, however, other modules in a CPPS cannot be fully taken into account at design time because they might not be known in advance. 
Furthermore, if a CPPS is modeled as an individual system and its modules as subsystems, the static separation between system and context required in systems engineering is not possible. Modules from the context of a CPPS become part of it during collaboration, and modules leaving the CPPS become part of the context. \cite{CrestBuchChapterFct}

Depending on the domain, the understanding of the concept of function varies \cite{inproceedingsEisenbart}. In this contribution, we understand a function as an abstract description of the purpose and behavior of a system. A function can be described by a combination of verb and noun at an early stage of the design of the system and has inputs and outputs \cite{VDI2221}.

\smallskip
On the one hand, MBSE is used to tackle complexity in functionalities during early engineering stages. On the other hand, when implementing system functions, there is ongoing research to make automation software more reusable and to offer functions in the sense of services in a service-oriented architecture. 
Recent approaches aiming to achieve a high degree of adaptability focus on modular manufacturing resources where each resource provides a set of functionalities as so-called \textit{skills}. 
Skills are seen as implementations of functions with a machine-readable interface description which can be invoked from a superordinate system such as a MES. 
For such descriptions, research approaches to skill descriptions make use of different technologies such as AutomationML \cite{DFL+_AnAutomationMLmodelfor_2017}, OPC UA information models \cite{ZAB+_SkillbasedEngineeringandControl_2019} or ontologies \cite{kocher_formal_2020}.

\smallskip
What is currently missing is a link between function models of MBSE on one side and the implementation and usage of functions in skill-based manufacturing on the other side. With this contribution, we want to create this link by answering the following research questions:
\begin{itemize}
    \item \emph{How can skill information which is necessary for implementation of system functions be connected to an abstract function model which is created during system design?}
    \item \emph{How can this function model be used to generate code for skills?}
\end{itemize}

Our goal is to achieve a well-defined method to document the development and distribution of functions in early engineering phases and at the same time ensure a high degree of adaptability of the plant at runtime through skills.

\smallskip  
After this introduction into the two research topics \emph{MBSE of system functions} and \emph{skill-based manufacturing}, the remaining sections of this contribution will continue on these two topics. In Section~\ref{sec:relatedWork}, related approaches for both these areas of research are discussed. Afterwards in Section~\ref{sec:ourApproach}, we present our approach which is a combination of model-based engineering and a skill-based implementation of functions. An evaluation of this approach is presented in Section~\ref{sec:evaluation} before this contribution ends with a summary and an outlook in Section~\ref{sec:summary}. 

\section{Related Work}
\label{sec:relatedWork}
\subsection{Approaches to Modeling System Functions}
Function-centric engineering approaches can be found in various domains\cite{SPESXT} and address a solution-neutral description of a system's expected behavior without specifying without a concrete binding to certain components \cite{VDI2221}. 
Functions are modeled in the early phase of engineering in order to facilitate the transition from informally documented requirements against a system to technical solution elements in the form of components. This is achieved via a step-by-step subdivision from initially abstract functions (called blackbox functions) at a high level of abstraction  to detailed descriptions of individual functional contributions (called white box functions). \cite{SPES2020}

A system model focusing on functions is also called a \emph{functional architecture} \cite{LaWe10}. 
Further advantages of describing systems by functional architectures involve the fact that they keep their validity if individual components of the system are replaced later by newer ones, as long as the replacement does not result in functional changes. 
In addition, functional architectures can ease system understanding for stakeholders with different know-how due to their simple form of representation.

A function-centric approach is thus particularly suitable for modeling dynamic CPPS, since specific details about components of the systems involved in such CPPS do not have to be known in order to describe them in terms of their inputs and outputs. Since from the point of view of a single module vendor, the actual specifications of other modules are typically not known, the omission of descriptions of specific components is even necessary when modeling CPPS.\cite{CrestBuchChapterFct}

After identifying in \cite{RequirementsPublication} what particular content should emerge from a function-centric, dynamic model of a system group, the Systems Modeling Language (SysML) was chosen as the appropriate modeling language. Since SysML itself does not consider functions, an extension of the language by a profile was necessary. In addition to this extension, a methodology for its use is also needed to guide modeling of system group model.\cite{hayward.approach.2022}

There are no SysML-based approaches that combine methods of function-centered engineering and modeling dynamic system groups yet.
Furthermore, existing approaches of using SysML only consider individual systems and their static relationship to the context. Among others, \cite{LaWe10,Friedenthal.2012,Weilkiens.2020,Aleksandraviciene.2018, Gausemeier.2009,Kernschmidt.2019, Meissner.2014} describe approaches on how SysML should be used in different domains. 
Typically a top-down procedure is recommended, where requirements are modeled first (e.g. by using requirement diagrams ), followed by the alternating description of structural elements and their relationships (via block definition diagrams ), connections of these structural elements (via internal block diagrams and parametric diagrams) and their behavior (via use case diagrams, state machine diagrams, activity diagrams and sequence diagrams). 
The approaches differ in terms of the suggested order of the diagrams to be used. 
Most of the approaches address modeling of functional hierarchies to reflect different granularity levels from a \emph{blackbox} to a \emph{whitebox} perspective. 
Additionally, in \cite{Weilkiens.2020} typing of interfaces is discussed which is used to consider the compatibility of different inputs and outputs of functions. 
This can be used to determine the compatibility of content to be exchanged between collaborating modules. Furthermore, since functions of a system group must be derived from goals to be achieved, goals must be modeled in SysML. Weilkiens presents goal modeling approaches in SysML for this purpose and shows how goals can be related to function models \cite{Weilkiens.2020}.

\subsection{Approaches of Skill-based Engineering }
The term \emph{skill} describes an encapsulated automation function that provides a machine-readable interface description to increase adaptability of production systems. Skills have been studied in research for several years and are often compared to service-oriented architectures in information technology \cite{FeLo_Configurationmodelforevolvable_2012}.

The approach presented in \cite{PSA+_Plug&producebymodelling_2015} tries to increase flexibility of production assets by a machine-readable description of asset functions. This description is an extension of the classic views of \emph{product}, \emph{process} and \emph{resource} to include \emph{skills}, which are seen as a more high-level description that is decomposed into actions for execution.

In \cite{ZAB+_SkillbasedEngineeringandControl_2019}, an executable skill meta model in the form of an OPC UA information model is presented. The skill model is used to decouple the invocation interface of skills from their concrete implementation. An evaluation is shown in which interaction between different stations - implemented using IEC 61131 and IEC 61499 - was achieved.

A semantic skill model in the form of an ontology that is based on different industry standards is introduced in \cite{kocher_formal_2020}. It extends the concept of \cite{ZAB+_SkillbasedEngineeringandControl_2019} because skills described with this model can be executed using both OPC UA and web services. A skill-based manufacturing execution system \emph{SkillMEx} is presented which allows registering and executing single skills as well as complete production processes consisting of multiple skills.

Despite a large amount of contributions in the field of \emph{skills}, the question of how to combine a model-based engineering approach of machines with the development of control code in the form of skills has not received sufficient attention yet.

An approach to introduce model-driven development into PLC code development is presented in \cite{TSU+_Towardsindustrialapplicationof_2014}. The approach uses UML and generates \emph{complete}, (i.e., executable) code and even allows for debugging. However, as the generated code is rather monolithic, skills are not considered and no machine-readable interface description (e.g., in the form of AutomationML or an ontology) is generated.

Spitzer et al. present an approach to generate skills together with a corresponding simulation model \cite{SFS_ATLASAGenericFramework_2021}. The approach generates a manufacturer-independent PLC code representation before deploying it to a specific PLC. By using mechanical, behavior and material flow models, a simulation is simultaneously generated. While the approach's goal is similar to ours, there are a variety of differences: It is not based on MBSE, skills are represented using an OPC UA information model and PLC programming languages are considered as target systems.

Creating large models manually requires specialized knowledge and high efforts -- especially for ontologies. Thus, an automated method to generate the ontological skill model of \cite{kocher_formal_2020} was presented in \cite{KHC+_AutomatingtheDevelopmentof_2020}. While this method takes engineering artifacts such as 3D models of machines into account, it is, however, based on a classical, i.e. non model-based engineering workflow. Therefore, it doesn't provide a model-based connection between design and implementation information. This current contribution can be seen as an extension of \cite{KHC+_AutomatingtheDevelopmentof_2020} to ensure a model-based development of functions as skills.

\section{Engineering CPPS Functions using MBSE and Skills}
\label{sec:ourApproach}

\begin{figure*}[htb]
    \centering
    \includegraphics[width=\textwidth]{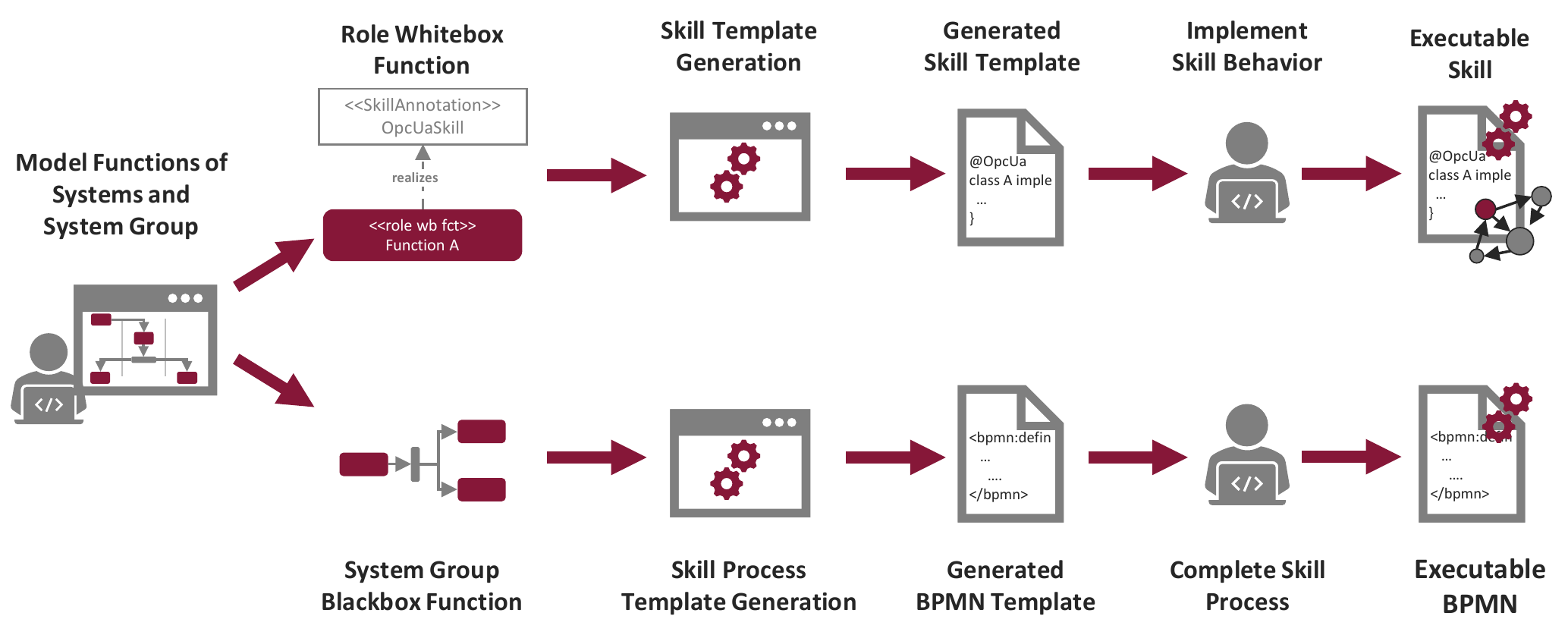}
    \caption{Overview of the two ways to generate executable code from an MBSE function model in our method. Starting from an MBSE function model of systems and system groups, both skill templates and skill process templates can be generated.}
    \label{fig:methodOverview}
\end{figure*}

In this Section, our MBSE method to modeling functions as well as our approach to generate code from such functions is presented. Figure~\ref{fig:methodOverview} depicts a rough overview over this method. After a user models both individual functions of a systems under consideration as well as aggregated functions of a system group, both individual and aggregated functions are used in order to generate code. 
The following two subsections introduce the MBSE method first before explaining the approach to code generation in the second subsection.

\subsection{A SysML Approach to Modeling Functions}
\label{sec:method_functionModeling}

In order to have a suitable basis for the implementation of skills, a model of the CPPS to which the individual modules should contribute is required. Based on the results of the project CrESt \footnotemark{} an approach was developed, which contains six steps to create a function-centered system group model \cite{hayward.approach.2022}. In the system group model, goals to be achieved, higher-level functions of the CPPS, individual functions for realizing the higher-level functions, potential roles to be assumed by modules in the CPPS and states of the roles for mapping functional dependencies are modeled and linked to each other. The six steps for building a CPPS model will be briefly described in the following subsections.
\footnotetext{https://crest.in.tum.de/}

\subsubsection{Modeling of Goals}
In the first step of the approach, the goals to be achieved by the group are modeled in order to obtain a suitable basis for deriving functions which should describe the solution-neutral functionalities of the group and its participants.

\subsubsection{Modeling of SG Blackbox Functions}
In the second step, black box functions are derived on the basis of the goals. Black box functions are intended to describe the externally observable behavior of the system group without specifying which individual systems must contribute to the realization of this behavior. 
Since these black box functions are assigned to the system group, they are also referred to as system group (SG) black box functions. 
As SysML distinguishes between structural and behavioral elements and the SG blackbox functions are behavioral specifications, an initial structural element of the system group is created in parallel to this step, to which the functions are assigned as owned behavior. 
The goals that have been modeled within the first step are linked to the modeled SG blackbox functions at the end of the second step to ensure that all goals of the group are considered by SG blackbox functions.

\subsubsection{Deriving Roles and Whitebox Functions}
The third step is the actual specification of the behavior of the SG blackbox functions at whitebox level. 
In this step, an activity diagram is created for each SG blackbox function, and whitebox functions are modeled and interconnected in it. 
Whitebox functions can be connected to each other either by control flows to define an execution sequence or by object flows to exchange content to be processed between them via inputs and outputs. 
Furthermore, in this step, whitebox functions are assigned to structural placeholders which should later be taken by collaborating CPPSs that are responsible to execute those whitebox functions. 
These placeholders are also referred to as \emph{roles} and are added to the SG element created in the second step by composition relations. 
Modeling the roles as a structure element allows them to be represented as allocated activity partitions in the activity diagrams of the SG black box functions to place the white box functions in them and thus assign them to the roles. 
Just as SG blackbox functions were assigned to the composite system element in the second step, this third step assigns whitebox functions to roles as owned behavior. 
Whitebox functions in the activity diagrams are therefore also referred to as \emph{role whitebox functions}. At the end of the third step, the system group model created up to this point shows which roles can be occupied within a system group and which role whitebox functions are needed to contribute to the realization of which SG blackbox functions.

\subsubsection{Describing Cross-Role Ports}
Since the behavior of the individual roles was specified in the third step and this includes an object exchange between roles, it is necessary to describe interfaces of the roles via which the cross-role inputs and outputs of functional content can be exchanged. For this purpose, all points at which cross-role object exchange takes place are identified using the activity diagrams of the third step. For each type of exchanged content, a separate port is then added to the structure element of the respective role.

\subsubsection{Linking Whitebox Function In- and Outputs to Cross-Role Ports}
In order to use the ports created in step 4, a change in the description of the behavior from the third step is required. The connections between role whitebox functions belonging to different roles are replaced by elements for sending or receiving signals. These signals can then be linked to ports created in the fourth step, thus realizing a functional exchange between multiple roles.

\subsubsection{Modeling and Aggregating Role States}
In the last step, a description of individual states of roles finally takes place in order to consider state-based functional dependencies. 
Role whitebox functions created in step 4 can be directly linked to modeled states in state diagrams. 
In addition, further interfaces can be used to exchange signals between roles that trigger state transitions and can thus be aggregated into system group states. State-level exchanges between roles are performed by so-called collaboration functions.

After the sixth step, a comprehensive system group model is available, which specifies the contributions of several future collaborating CPPSs by a function-centric description in order to enable the achievement of higher-level goals by coordinated behavior. During modeling iterative feedback is possible whenever it seems necessary.

\subsection{Code Generation for Skills}
After modeling functionalities of systems and system groups using the approach presented in the previous subsection, we aim to automatically generate code in the form of \emph{skills} for these functionalities.
We distinguish between the generation of individual skills of a system and the generation of a superordinate orchestration sequence which consists of individual skills and is later performed collaboratively by the system group. These two steps of our method will be presented in the following two subsections.

\medskip
\subsubsection{Generation of Individual Skill Templates}
After modeling functions of a system as described in Section~\ref{sec:method_functionModeling}, these functions shall be implemented as \emph{skills}. We use \mbox{SkilllUp}\footnotemark{}, a Java framework for simplified development of skills which was initially presented in \cite{KHC+_AutomatingtheDevelopmentof_2020}. With SkillUp, a developer only has to program the actual skill behavior - a task that would also be required with a conventional control approach. The necessary state machine, an OPC UA or web server to call the skills, and the mandatory ontological description are created automatically.
For this purpose, a Java class has to be created, which needs to be tagged with a couple of Java annotations. These annotations are evaluated as soon as a skill is deployed in a special runtime to register skills with a higher-level system and to interact with them. Currently, such a class with annotations needs to be created manually which still requires some experience, may lead to errors and lacks connection to MBSE design models. 

\footnotetext{https://github.com/aljoshakoecher/skill-up}

An example of a \emph{skill class} that can be used to add two numbers can be seen in Listing~\ref{lst:exampleSkill} and the most relevant annotations are briefly introduced as follows: 
\verb|@Skill| is a required annotation that marks a class as a skill. It is used to specify the automated generation of the skill interface (currently either OPC UA or webservices) as well as the containment of the skill by a certain module. The SkillUp runtime looks for classes annotated with \verb|@Skill| to register new skills as soon as they are deployed.

\verb|@SkillParameter| and \verb|@SkillOutput| can be used to specify which parameters can be set and which result values may be retrieved from outside the skill, respectively. The annotation \verb|@StateMachine| provides access to the internal state machine of a skill which is generated automatically. The provided instance can be used to control the state machine from within a skill implementation, e.g., abort it in case of certain conditions. 
Every skill adheres to this state machine, so the actual skill behavior must be assigned to the different states (e.g., \emph{starting}, \emph{execute}, \emph{aborting}). Thus, every state's behavior is implemented in its own method which is annotated with the name of the corresponding state. In the example of Listing~\ref{lst:exampleSkill}, the main processing shall be executed during the \emph{execute} state and is therefore implemented inside a method which carries a corresponding annotation.

\begin{lstlisting}[
    label={lst:exampleSkill},
    language={Java}, 
    caption={An example of the annotations used to generate a skill. This very simple skill allows to add two numbers and returns the sum. After deploying, it can be invoked via OPC UA. Please note that this is a simplified example and arbitrary actions can be implemented using SkillUp.}, 
]
@Skill(type=OpcUaSkillType.class, moduleIri="https://hsu-hh.de/modules#Module1")
public class AdditionSkill {

    @SkillParameter
    int a,b;

    @SkillOutput
    int result;		
    
    @StateMachine
    Isa88StateMachine stateMachine;
    
    @Execute	
    public void aPlusB() {	
        this.result = this.a + this.b;
    }
}
\end{lstlisting}

\medskip
With this publication, we now create a defined link between a model-based engineering approach for designing functions with a skill-based approach for implementing them. This is achieved through automated code generation of \emph{skill class templates}, i.e. Java class templates like the one in Listing~\ref{lst:exampleSkill} with all relevant annotations that only lack method implementations.

As a preparation for the code generation an additional class diagram, a so-called \emph{deployment diagram} which maintains separation of the engineering phase from the actual implementation, is created. Every role whitebox function to be implemented is added to this deployment diagram.

With an additional UML interface element stereotyped \emph{SkillAnnotation}, all information of the \verb|@Skill| annotation are captured. This element is treated as a regular interface that the activity implements, i.e., it is connected to the activity through a realization connection. We use \emph{Enterprise Architect} for modeling and code generation. Annotations of a UML class are automatically transferred to Java, but in order to transfer the annotation values from a \emph{SkillAnnotation} interface, the code generation capabilities of Enterprise Architect needed to be extended. A code generator extension was developed that checks implemented interfaces of a UML class for annotations and transfers them to the code generated for that UML class.

Furthermore, the annotation \verb|@StateMachine| is added with a corresponding variable automatically.  All input and output parameters of the activity are also transferred to members of the Java class and annotated with \verb|@SkillParameter| and \verb|@SkillOutput|, respectively. Furthermore, for every state of the state machine, a method annotated with the state name is generated. This results in a complete skill class template, so that a developer now only has to implement the methods needed. Unneeded ones may be left empty or removed from the class.
After all needed methods have been implemented and the skill is thus complete, it can be executed using the \emph{SkillUp} runtime. 

But a production processes most likely consists of more than one skill. The following subsection describes how a higher-level sequence coordinating the interaction of multiple skills is derived from the whitebox functions of a complete system group.

\medskip
\subsubsection{Generation of Skill Processes}

We have developed a concept to model and execute \emph{skill processes}, i.e., sequences of interacting skills, using \emph{Business Process Model and Notation} (BPMN) \cite{KVF_ModelingandExecutingProduction_2022}. BPMN is a modeling language originating from the field of administrative processes. It is defined as a comprehensive standard for modeling processes which shall be understood by all human process participants and which also defines formal execution semantics that allow for an automated execution of BPMN processes through so-called \emph{BPMN engines}. \cite{OMG_BusinessProcessModeland_2011}
Zor et al. have shown that BPMN can be used not only for business processes, but also in manufacturing to help companies become more flexible \cite{SKF_UsingBPMNforModeling_2010}.

In our concept to use BPMN for skill processes, we make use of \emph{ServiceTasks}, a BPMN element that allows to trigger a web service or another application via delegation to a custom class. As such a custom delegation class, we implemented a \emph{SkillExecutor} that is able to invoke single skills and is integrated into the \emph{Camunda}\footnotemark{} BPMN engine. For every skill in a BPMN process, the engine delegates all necessary information which are stored in a ServiceTask to the SkillExecutor. The SkillExecutor returns execution to the process engine as soon as it completes and throws BPMN errors that can be handled by the engine if a skill enters an error.

\footnotetext{https://camunda.com/}
\smallskip
Until now, these executable skill processes are modeled manually without any reference to plant engineering activities. With this contribution, we establish a link between an MBSE model of the functions of a system group with an executable process description composed of skills.
As presented in Section~\ref{sec:method_functionModeling}, activity diagrams are used to model system group functions.
Accordingly, an automated approach to transfer the elements of an activity diagram into a BPMN process has been developed. As both activity diagram and BPMN processes are similar in terms of the semantics of their elements, this approach is a rather direct mapping.

The developed BPMN code generator is a C\# algorithm with classes for all elements of a UML activity diagram. The algorithm is used as an add-in to \emph{Enterprise Architect}. Every class provides a functionality to get serialized into a BPMN compliant XML snippet. 
When invoking the code generator, for every element of a selected activity diagram, an instance of the corresponding C\# class is created. Afterwards, the whole diagram is serialized by serializing all elements and combining them to one BPMN process. This process can then be imported into our skill-based manufacturing platform \emph{SkillMEx}\footnotemark{} where further details such as additional tasks can be added manually.
Finally, parameter values for all skills need to be added before SkillMEx can be used to execute a BPMN process.

\footnotetext{https://github.com/aljoshakoecher/SkillMEx}

\section{Evaluation}
\label{sec:evaluation}

\begin{figure}[htb]
    \centering
    \includegraphics[width=0.8\linewidth]{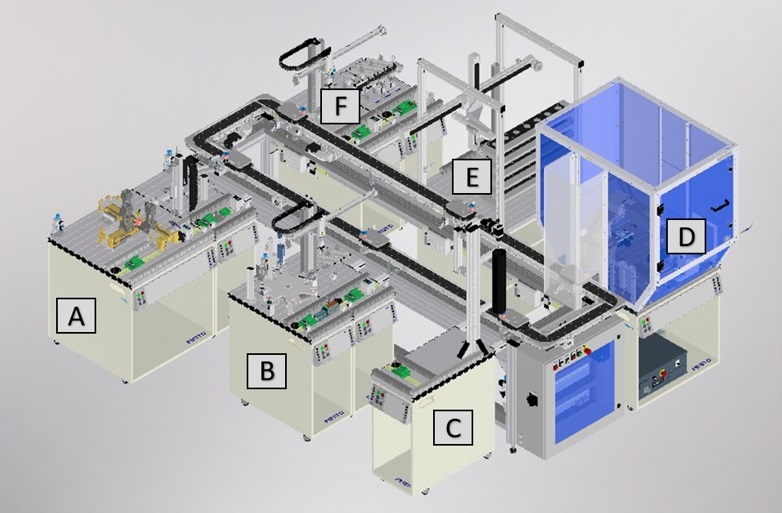}
    \caption{Physical model of a modular lab plant consisting of six modules.}
    \label{fig:MPS}
\end{figure}

As an evaluation use case, we consider a modular lab plant, which is located in the laboratories of the Institute of Automation at Helmut Schmidt University in Hamburg, Germany. 
The plant can be used to manufacture various products, including a thermometer, which is composed of a base body that must be handed out by a storage module, processed by a manufacturing module and checked by a quality control module.
The base body is assembled with a lid and finally stored in a storage module. The various modules of this plant can be exchanged via plug and produce and connected to the overall plant in a decentrally controlled and flexible manner. Figure~\ref{fig:MPS} shows the overall structure of the plant. The different modules A-F are connected to a central conveyor belt via which workpieces are transported between the modules on a carrier.

Using this plant, the approach for creating a system group model described in Section~\ref{sec:method_functionModeling} has been applied and will be presented here in abbreviated form. Due to the high complexity of the model, only excerpts are shown. The complete model is available online\footnotemark{}.

\footnotetext{https://github.com/hsu-aut/MPS-SystemGroupModel}

Within the first three steps of the approach, goals of the CPPS, its System Group (SG) blackbox functions to achieve these goals and role whitebox functions to specify the individual contributions to the realization of the SG blackbox functions and their assignment to structural elements (roles) are carried out.

\begin{figure}[htb]
    \centering
    \includegraphics[width=0.9\linewidth]{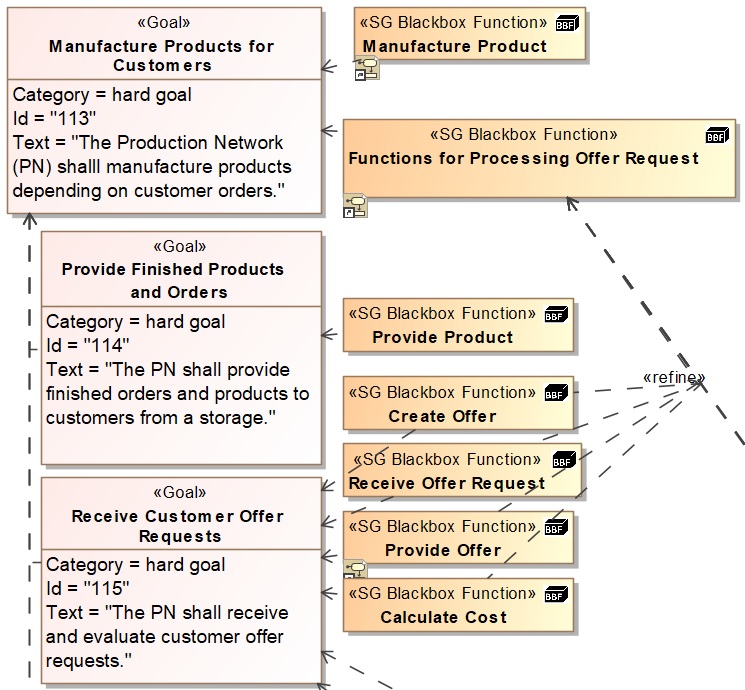}
    \caption{Excerpt of modeled goals to satisfy SG blackbox functions.}
    \label{fig:Goals}
\end{figure}

Figure~\ref{fig:Goals} shows an excerpt from the comparison of modeled goals and SG blackbox functions. The goal with ID 10 named \emph{Manufacture Products for Customers} is realized by the two SG blackbox functions \emph{Manufacture Product} and \emph{Functions for Processing Offer Request}, with additional SG blackbox functions specified for refinement such as \emph{Receive Offer Request} and \emph{Create Offer} for the goal of interacting with the customer.

\begin{figure}[htb]
    \centering
    \includegraphics[width=0.7\linewidth]{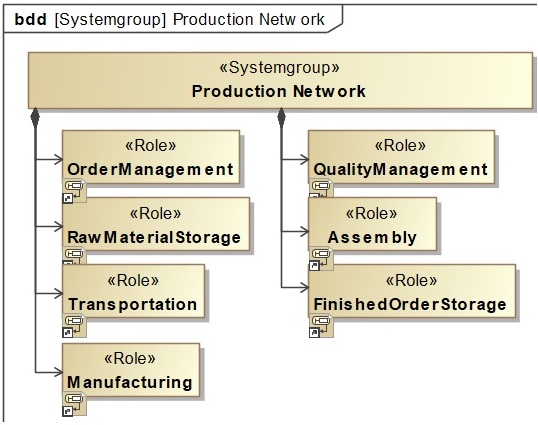}
    \caption{Structural elements of the system group and its related seven roles.}
    \label{fig:Roles}
\end{figure}

In order to implement the SG blackbox function \emph{Manufacture Product}, several role whitebox functions are required, which can be assigned to different roles to be occupied within the CPPS. An overview of the roles to be occupied can be seen in Figure~\ref{fig:Roles}. Based on the existing physical model of the plant in Figure~\ref{fig:MPS}, roles for order management, for the storage of raw materials, for the processing of raw materials by manufacturing functions, for quality inspection, for the assembly of components, for storing the finished products, and for transport between the modules are considered. Note that there are multiple modules that can assume a role.

A behavioral description of the SG blackbox function \emph{Manufacture Product} is done within an activity diagram, in which the different roles that should contribute to this function have been modeled as \emph{Allocated Activity Partitions}. The individual role whitebox functions have been added within their respective role areas for direct assignment.

\begin{figure*}[htb]
    \centering
    \includegraphics[width=0.9\textwidth]{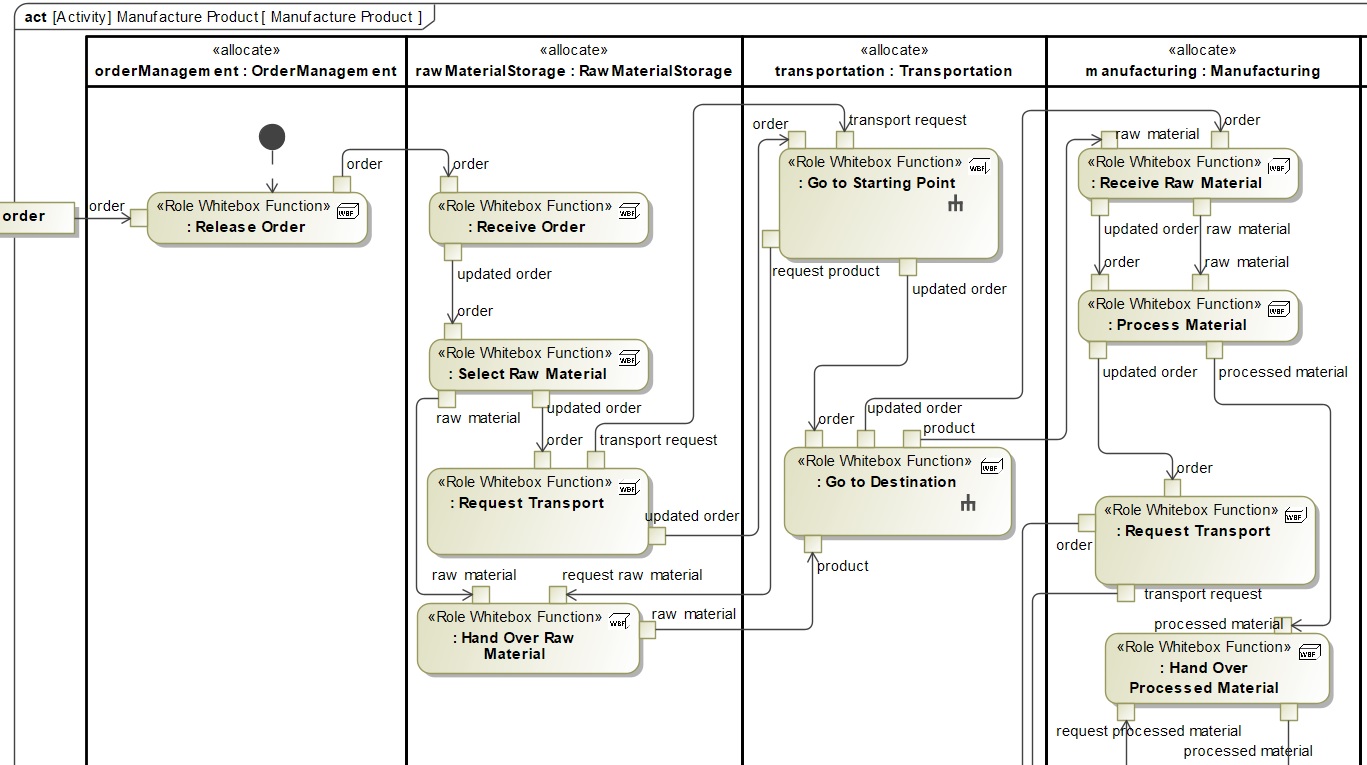}
    \caption{Excerpt from the whitebox specification of the SG blackbox function \emph{Manufacture Product} containing various role whitebox functions assigned to their roles.}
    \label{fig:ManufactureProduct}
\end{figure*}

This behavioral specification of the SG blackbox function \emph{Manufacture Product} is shown in Figure~\ref{fig:ManufactureProduct}, in which the order is released by the role \emph{OrderManagement}, so that the role \emph{RawMaterialStation} can then start handing out the required material. The selection of a required material is derived from the description of the order, which is exchanged between the connected role whitebox functions via their inputs and outputs. The role \emph{Transportation} then takes over the extracted raw material to transport it to the role \emph{Manufacturing}, so that it can be processed according to the specification of the order.

Finally, in steps 4 and 5 of the approach, the interfaces between the individual roles are specified so that they can perform the exchange of, among other things, \emph{order}, \emph{transport request} and \emph{raw material} as it can be seen in Figure~\ref{fig:ManufactureProduct} between the roles. In addition, a role-specific description of the behavior is modeled, in which these transitions between the roles are replaced by signals to link the behavior to the interfaces.

In the last step of the approach, the various role states are modeled in order to link them to the different role whitebox functions. This is intended to ensure that a module that assumes a role only executes the associated functions when it is in the appropriate state. 

\begin{figure}[htb]
    \centering
    \includegraphics[width=0.8\linewidth]{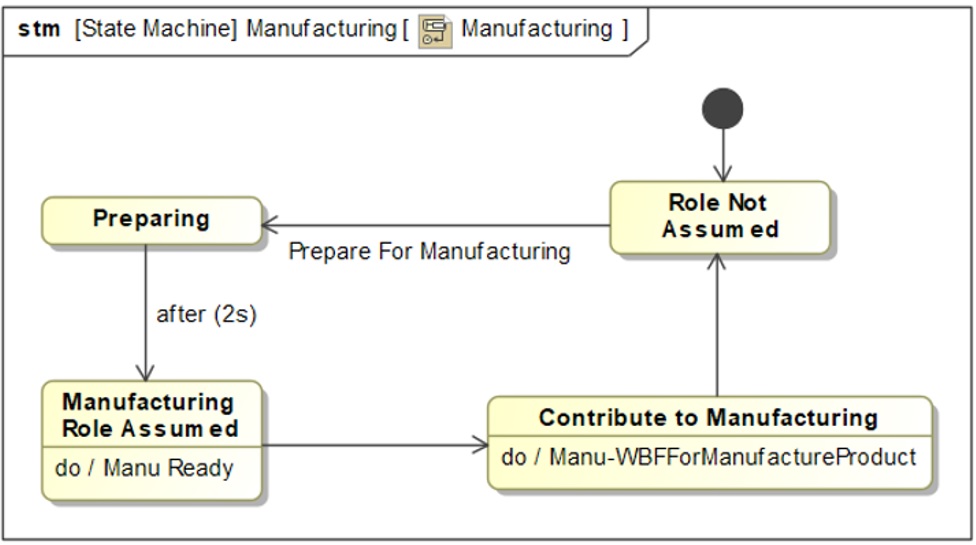}
    \caption{Exemplary states of the role \emph{Manufacturing} and linked role whitebox functions}
    \label{fig:States}
\end{figure}

An excerpt of the state description of the \emph{Manufacturing} role can be found in Figure~\ref{fig:States}. 
The lower right state \emph{Contribute to Manufacturing} in the Figure~\ref{fig:States} leads to an execution of the role whitebox function \emph{Manu-WBFForManufacturing}. This role whitebox function contains the collected contributions from the role \emph{Manufacturing} for the SG blackbox function \emph{Manufacture Product}. 
The other role states are modeled in the same way but can't be shown here due to limited space.

\smallskip
For this evaluation, only the roles \emph{RawMaterialStorage}, \emph{Transportation} and \emph{Manufacturing} are implemented using skills. For this purpose, their role whitebox functions are placed in a separate deployment diagram and connected to a \emph{SkillAnnotation} interface with all relevant skill information. For this evaluation, all skills were implemented with an OPC UA interface. For each function, a Java class template is generated that is subsequently completed with behavior. As the actual control logic of the plant already existed before in the form of IEC 61131 PLC code, all skills act as proxies that communicate with a skill-based manufacturing system and eventually trigger the actual PLC control logic. For this purpose, skills were deployed on additional Raspberry PIs at each module.

The sequence between the roles \emph{RawMaterialStorage}, \emph{Transportation} and \emph{Manufacturing} as modeled in Fig. \ref{fig:ManufactureProduct} was converted into a BPMN process that sequentially invokes the separate skills. Details of this process such as parameter values were added manually in order to execute the process.
As soon as an order is received (i.e., the BPMN process is started), material information is taken from the order and used by the storage module to select the according raw thermometer base. Then, a carrier is requested which in turn causes the conveyor to start. As soon as a carrier arrives, the raw material is moved to the carrier and the conveyor transports the carrier to the manufacturing module where it finally is processed.

With this evaluation, we were able to show that manual effort is eliminated when creating both single skill classes and skill processes. Furthermore, automated skill code generation leads to a reduction of errors caused by manual interaction (e.g., typing mistakes). In addition, the presented model-based approach ensures a structured transition from the design to the implementation phase and links the relevant information in a shared model.

\textbf{}

\section{Summary \& Outlook}
\label{sec:summary}

This contribution presented an approach to link model-based system engineering activities of system functions with an implementation based on skills. With this approach, it is possible to connect implementation information with information from earlier engineering phases (e.g., requirements engineering and functional architectures).

In order to transfer information into the implementation phase, code templates are automatically generated for both individual system functions and for role whitebox functions of system groups. While individual functions are transformed to Java class templates that can be used inside a skill runtime, role whitebox functions are transformed to executable BPMN processes that can be used to orchestrate multiple skills. The algorithm to generate BPMN processes from UML activity diagrams is publicly available.\footnotemark{}
\footnotetext{https://github.com/aljoshakoecher/BPMN-Code-Generator}

Currently, the presented approach only covers skill information although the ontology published in \cite{kocher_formal_2020} is able to hold information about the structure of machines, too. In future research, this information could also be transformed from an MBSE artifact, e.g., from SysML block definition diagrams.

\bibliographystyle{IEEEtran}
\bibliography{references} 

\end{document}